\documentclass[apj]{emulateapj}
\usepackage{natbib}
\usepackage[caption=false]{subfig}
\usepackage{graphicx}
\usepackage{makecell}
\usepackage{color}
\shorttitle{}
\shortauthors{Dunn et al.}

\begin{document}

\title{the role of gravitational recoil in the assembly of massive black hole seeds}

\author{Glenna Dunn\altaffilmark{1},  Kelly Holley-Bockelmann\altaffilmark{1,2}, Jillian Bellovary\altaffilmark{3,4,5}}
\affil{1 Vanderbilt University, Nashville, TN, USA}
\affil{2 Fisk University, Nashville, TN, USA}
\affil{3 Queensborough Community College, New York, NY, USA}
\affil{4 American Museum of Natural History, New York, NY, USA}
\affil{5 CUNY Graduate Center, New York, NY, USA}
\email{glenna.dunn@vanderbilt.edu}

\begin{abstract}
When two black holes merge, the asymmetric emission of gravitational waves provides an impulse to the merged system; this gravitational wave recoil velocity can be up to 4000 km s$^{-1}$, easily fast enough for the black hole to escape its host galaxy.  We combine semi-analytic modeling with cosmological zoom-in simulations of a Milky Way-type galaxy to investigate the role of black hole spin and gravitational recoil in the epoch of massive black hole seeding.  We sample four different spin distributions (random, aligned, anti-aligned, and zero spin),  and compare the resulting merger rates, occupation fractions, and MBH-host relations with what is expected by excluding the effect of recoil.  The inclusion of gravitational recoil and MBH spin in the assembly of MBH seeds can reduce the final $z=5$ MBH mass by up to an order of magnitude.  The MBH occupation fraction, however, remains effectively unaltered due to episodes of black hole formation following a recoil event.  While electromagnetic detections of these events are unlikely, LISA is ideally suited to detect gravitational wave signals from such events.

\end{abstract}


%%%%%%%%%%%%%%%%%%%%%%
\section{Introduction}
%%%%%%%%%%%%%%%%%%%%%%

%\comment{MBH BACKGROUND}
By now, massive black holes (MBH) are a well-established galaxy component found in an astonishing variety of galaxy hosts, from bulgeless spirals \citep{Satyapal09,ArayaSalvo12,Simmons13,Simmons17}, low-surface brightness galaxies \citep{Subramanian16}, and dwarfs \citep{Reines11, Baldassare18, Secrest15, Marleu17,Mezcua18} to the brightest cluster ellipticals \citep{McConnell12,FerreMateu15}, and at a wide range of redshifts \citep{Banados18}.  These MBHs must have formed within the first billion years of the Universe and likely in the form of massive `seed' black holes \citep{Haiman01} via the remnants of Population III stars \citep{Madau01,Johnson07,Xu13} (see, however \citet{Smith18}) or direct collapse \citep{Loeb94,Koushiappas04,Begelman06,Lodato06,Haiman06,Spaans06}.

Observations of the local Universe describe a correlation between the mass of the central black hole and the velocity dispersion of the bulge component of the host galaxy, commonly known as the $M_{\rm BH}-\sigma$ relation \citep{Gebhardt00,Ferrarese00,Tremaine02,Gultekin09,McConnell13,Kormendy13,Woo13,Saglia16}.  Related MBH-host scaling relations abound, including $M_{\rm BH}-M_{\rm bulge}$, $M_{\rm BH}-L_{\rm bulge}$, $M_{\rm BH}-M_{\rm stellar}$, and  $M_{\rm BH}-M_{\rm halo}$ \citep[see e.g.,][]{Kormendy13}.  An MBH and its host halo co-evolve as a system, and there is a wealth of literature exploring how this relationship can be harnessed to understand galaxy formation \citep{DiMatteo08,Micic07,Barausse12,Antonini15} and MBH demographics \citep[e.g.,][]{Merritt01,Micic08,Volonteri09,Heckman14,Tremmel17}.

Any MBH-host co-evolutionary model, however, must contend with MBH binary dynamics, such as three-body scattering, gravitational wave emission, and gravitational wave recoil \citep{Merritt05}.  Gravitational recoil is caused by the anisotropic emission of gravitational radiation during a merger of two compact objects \citep{Bekenstein73}.  As MBHs merge, asymmetries in the merger configuration cause an asymmetric gravitational wave emission pattern. This, in turn, radiates linear and angular momentum in a preferred direction, which imparts a gravitational kick to the merged object to conserve momentum.  This kick can range up to 5000 km s$^{-1}$ \citep{Campanelli07b, Lousto11}, with the high velocity `superkick' tail caused by spins anti-aligned to one another, perpendicular to the orbital plane, and highly eccentric \citep{Gonzalez07,Campanelli07b,Hermann07,Baker08}, and even larger kicks occurring for partially aligned spins in certain `hangup' configurations \citep{Lousto11}. More typical recoil velocities are of order a hundred km s$^{-1}$ for circular orbits, low spins, and aligned spin configurations \citep{Campanelli07a}.

%\comment{BINARIES ESCAPING HOST}

The magnitude of these kick velocities presents a challenge to our simple picture of central black hole evolution.  Low mass halos and globular clusters, for example, cannot retain seed MBHs suffering even mild kicks \citep{Volonteri07,HolleyBockelmann08}.  Indeed, recoil velocities likely exceed the escape velocities of all but the most massive host halos, ejecting the new MBH from its host \citep{Micic06, Micic11, Sijacki09, Gerosa15,Blecha16, Kelley17}.   Even moderate kicks can eject a MBH from high$-z$ galaxies, when halo virial masses are still small and halos lack sufficiently deep potential wells to retain these merged objects {\citep{Merritt04,Micic11,Schnittman07a,Volonteri07}.  Gravitational recoil may be responsible for a population of wandering black holes in the halos of galaxies \citep{Libeskind06,Micic11}, though these are also expected from long dynamical friction sinking timescales \citep{Bellovary10, Micic11}.  MBHs kicked at velocities that approach, but do not exceed, the escape velocity of the host, may oscillate on large orbits unless the host is gas rich, in which case gas exerts drag on the MBH and may restrict its orbit to the central region of the host galaxy \citep{Blecha11,Guedes11}.

%\comment{EFFECTS OF RECOIL ON GALAXY PHYSICS}
Recoil events are likely to be suppressed in gas-rich environments because accretion can force the spins of an MBH binary to align \citep{Bogdanovic07,Barausse12}, resulting in minimal kicks \citep{Dotti10}.  If MBH-MBH mergers occur as a result of galactic mergers, torque due to gas accretion may force the MBH spin axes to align with that of the galactic disks, leading to lower recoil velocities and consequently a retention of MBHs at the centers of host galaxies after the merger \citep{Bogdanovic07,Blecha16}.  Though circumbinary gaseous disks may provide the conditions necessary to align the spins of two merging black holes via the Bardeen-Petterson effect, rapidly spinning black holes may not have sufficient time to align, and therefore remain subject to large recoil velocities \citep{Lodato13}.

By displacing the MBH, gravitational wave recoil can significantly alter the co-evolution of an MBH and its host.  Outside the gas-rich center of a galaxy, the growth of a displaced MBH is stifled and the resulting low mass accretion interrupts the AGN feedback process \citep{Blecha08, Sijacki09}.  Without a central MBH to fuel, the gas can instead form stars and build up a bulge \citep{Blecha11}, which can have profound effects on MBH scaling relations.  When gravitational recoil is considered, the nuclear black hole occupation fraction decreases with decreasing stellar mass regardless of both morphology and redshift, though this effect is particulary significant in pseudo-bulge galaxies that experience less mergers, and therefore have less opportunities to refill the bulge with another MBH \citep{IzquierdoVillalba20}.  Gravitational recoil also alters the black hole mass-bulge velocity dispersion relation  by increasing the bulge velocity dispersion \citep{Blecha11} and increases the scatter in the black hole mass-bulge mass relation for brightest cluster galaxies \citep{Gerosa15}.  Gravitational recoil is an exigent component of any study of MBH-host galaxy co-evolution.

%\comment{PURPOSE}
The purpose of this work is to study the effect of gravitational recoil due to MBH-MBH mergers during the early assembly of MBH seeds.  We follow MBH seeds formed via direct collapse \citep{Dunn18} as they co-evolve with their host halos and merge with other MBHs.  In this paper, we investigate how gravitational recoil affects the high redshift MBH mass function, merger rate, occupation fraction and MBH-host scaling relations. Since MBH seeds are thought to be sown in the pre-reionization era, it may seem that the impact of recoil would be all but impossible to observe. However, the gravitational waves generated from MBH mergers are the loudest known source in the Universe and easily penetrate through matter, opening a gravitational window onto the cosmic dawn.  The Laser Interferometer Space Antenna (LISA), for example, is specifically designed to detect seed black hole mergers that may spawn the first quasars, translating into an observational requirement to detect the coalescence of MBHs in the mass range of $10^3-10^5 M_{\odot}$ between redshifts 10$-$15 and $10^4-10^6 M_{\odot}$ at redshifts greater than 9, with signal-to-noise ratios in the hundreds \citep{LISA17}.  With the advent of LISA, gravitational recoil kicks larger than $\sim$ 500 km s$^{-1}$ may even be directly observable through a Doppler shift of the gravitational wave signal during ringdown \citep{Gerosa16}.  The proposed X-ray flagship observatory, Lynx, is designed  with high throughput and fine angular resolution to observe highly accreting seed black holes at such high redshifts, as well. The synergy between LISA and Lynx could give us the opportunity to construct MBH seed demographics and follow complementary channels of seed growth \citep{Colpi19}.

%\comment{SUMMARY}
The paper is organized as follows: in Section \ref{Sims}, we describe the simulations used to generate the MBH merger tree; in Section \ref{Model} we discuss the semi-analytic gravitational recoil model; and in Section \ref{Results}, we discuss the importance of incorporating gravitational recoil into models that make predictions about the assembly of MBH seeds and the relationships between MBHs and their host galaxies.

%%%%%%%%%%%%%%%%%%%%%%%%%%%%%%%%%%
\section{Simulations} \label{Sims}
%%%%%%%%%%%%%%%%%%%%%%%%%%%%%%%%%%
Our study is based on simulations using the N-body+Smooth Particle Hydrodynamics (SPH) tree code \textsc{Gasoline} \citep{Stadel01,Wadsley04,Wadsley17}, which study the evolution of MBH seeds in cosmological zoom-in simulations.  We simulated the formation of a redshift zero Milky Way-mass halo until $z=5$, employing physically motivated prescriptions for black hole formation, accretion, feedback and mergers \citep{Bellovary10,Dunn18}.  MBHs in these simulations form from dense, converging, low-metallicity gas particles experiencing a Lyman-Werner specific intensity above a criticial threshold, $J_{\rm crit}$, which we vary from $30$ to $10^3 J_{21}$.  When an MBH forms, it subsumes the mass of its parent gas particle, forming with a mass of $2\times 10^4 M_{\odot}$.  For a detailed description of the simulations that generated the data used in this paper, see Section 2 of \citet{Dunn18}.

One of the major findings from our previous work was that multiple MBH seeds can form in a single halo in sequential bursts.  These multiples form despite concerted efforts to prevent spurious MBH overproduction.  No two MBH seeds are permitted to form at the same time step within two softening lengths of each other.  If more than one gas particle meets the MBH formation criteria within this volume at the same time step, only the most bound particle becomes an MBH seed, and the remaining MBH candidates revert to their parent gas particles. Even with this restriction, the multiplicity of MBH seeds forming within small volumes of space and windows of time leads to MBH-MBH mergers.  We repurpose the $J_{\rm crit} = 10^3 J_{21}$ simulation here to investigate the role of mergers in MBH assembly.  Of the three simulations in this suite, this $J_{\rm crit}$ invokes the most strict MBH formation recipe, resulting in the lowest MBH formation efficiency.

We note here that the possibility of forming more that one direct collapse black hole seed per halo is controversial.  One strong impediment to multiple seeds is the rate of gas inflow required by direct collapse, which must exceed $0.1 M_{\odot}/\rm yr$ for $\sim1-10$ Myr \citep{Hosokawa13,Alexander14,Umeda16}.  It may be difficult for this rapid collapse to occur at multiple locations or multiple occasions in a single halo.  Additionally, the separations between a direct collapse host halo and a neighboring starforming halo may need to be finely tuned to provide the necessary Lyman-Werner radiation without subjecting the MBH-forming halo to tidal disruptions \citep{Chon16}.  However, this work, like all simulations, is limited by its resolution, and MBH and star formation in \textsc{Gasoline} simulations are designed to rely only on local gas physics.  Furthermore, since the Lyman-Werner sources in \citet{Dunn18} are nearly exclusively {\em internal} to the the direct collapse host halo, this configuration may allow Lyman-Werner radiation to reach the direct collapse site without the threat of tidal disruption.

%%%%%%%%%%%%%%%%%%%%%%%%%%%%%%%%%%
\section{Semi-analytic Model} \label{Model}
%%%%%%%%%%%%%%%%%%%%%%%%%%%%%%%%%%
In {\textsc Gasoline}, MBH seeds can grow through mergers that occur when two sink particles are less than two softening lengths apart and  their relative velocities are small: $\frac{1}{2} \Delta \vec{v}^2 < \Delta \vec{a} \cdot \Delta \vec{r}$, where $\Delta \vec{v}$ and $\Delta \vec{a}$ are the differences in velocity and acceleration of the two black holes, and $\Delta \vec{r}$ is their separation.  Our \textsc{Gasoline} simulations do not include gravitational recoil due to MBH mergers.  To incorporate gravitational recoil into the simulations in post-processing, we follow the work of \citet{Schnittman07a,Campanelli07a} to construct a semi-analytic model of gravitational recoil events based on MBH interactions in the simulation.  The empirical expression for the total sum of recoil velocities described in \citet{Campanelli07a} is based on the post-Newtonian expression for linear momentum lost to gravitational radiation during a merger of spinning, unequal mass black holes.  The authors demonstrate that recoil velocities are dominated by the spin contribution, and construct an empirical formula that sums the contributions of mass and parallel and perpendicular angular momentum components.  This heuristic formula depends on the mass ratio $q$, the specific spin magnitudes $\alpha_1$ and $\alpha_2$, and the angle $\Theta$ between the in-plane spin component and the infall direction \citep{Campanelli07a}.

We construct four MBH merger history models to compare with the `no kick' model from our original simulation: no spin ($\alpha_1=\alpha_2=0$), random spins ($-1 < \alpha_{1,2} <1$; $0 < \Theta < \pi$; $\alpha_1 \neq \alpha_2$), aligned spins ($\alpha_1 \neq \alpha_2$; $\Theta=0$), and anti-aligned spins ($\alpha_1 \neq \alpha_2$; $\Theta=\pi$).  The mass ratios used in these models are drawn from the fiducial simulation, while the spin amplitudes and orientations are drawn from a random distribution. The orientations of the aligned and anti-aligned spins are in the plane of the binary orbit.  We run 100 Monte Carlo trials of each spin distribution model, and summarize some aspects of the MBH populations produced by these models in Table \ref{SimSummary}.

\begin{table*}
  \begin{center}
  \begin{tabular}{lcccccc} 
   \hline
   Recoil model & $N_{\rm mergers}$ & $N_{\rm ejections}$ & \thead{max($M_{BH}/M_{\odot}$)\\ at $z=5$} & \thead{Q$_1$($M_{BH}/M_{\odot}$)\\ at $z=5$} & \thead{Q$_3$($M_{BH}/M_{\odot}$)\\ at $z=5$}\\ 
  \hline
   No kick            & 45 & 0  & $4 \times 10^5$   & $2.7\times10^4$ & $1.1\times10^5$\\
   No spin            & 35 & 15 & $1.1 \times 10^5$ & $2.7\times10^4$ & $5.4 \times10^4$\\
   Random spins       & 29 & 25 & $4.6 \times 10^4$ & $2.7\times10^4$ & $3.7\times 10^4$\\
   Aligned spins      & 35 & 14 & $1.1 \times 10^5$ & $2.7\times10^4$ & $3.9\times10^4$ \\
   Anti-aligned spins & 28 & 26 & $4.5 \times 10^4$ & $2.7\times10^4$ & $3\times10^4$ \\
  
  \hline
  \end{tabular}
  \caption{\small{\textsc{simulation parameters} Summary of selected results of the simulations presented in this paper.  (1)  Spin distribution model, (2)  Total number of mergers, (3)  Number of merged MBHs that are ejected from their host halo, (4)  Mass of the largest MBH at $z=5$, (5)  First quartile (25$^{th}$ percentile) of MBH masses at $z=5$, (6)  Third quartile (75$^{th}$ percentile) of MBH masses at $z=5$.  Values for the aligned, anti-aligned, and random spin distributions are averaged results of 100 iterations of the merger history.}}
  \label{SimSummary}\end{center}
  \end{table*}

%%%%%%%%%%%%%%%%%%%%%%%%%%%%%%%%%%%%%%%%%%%%%%%%%%%%%%%%%%%%%%%%%%%%%%%%%%%%%%%%%%%%%%%%%%%%%%%%%%%

%%%%%%%%%%%%%%%%%%%%%%%%%%%%%%%%%
\section{Results} \label{Results}
%%%%%%%%%%%%%%%%%%%%%%%%%%%%%%%%%%%%%%%%%%%%%%
%\subsection{Subsection}
%%%%%%%%%%%%%%%%%%%%%%%%%%%%%%%%%%%%%%%%%%%%%%
All of the recoil models discussed in this work change the numbers of mergers, ejections, and $z=5$ final MBH masses of the black holes produced in these simulations (Table \ref{SimSummary}).  The random spin and anti-aligned spin recoil models yield the most drastic difference in the $z=5$ MBH populations, as they generate larger recoil velocities than other recoil models.  In this section, we discuss some of ways that the incorporation of gravitational recoil events into our simulations changes the predicted MBH population.
%% FIGURE 1 -- OCC FRAC

The occupation fraction of MBHs likely encodes information about the details of MBH formation and early assembly.  We define the MBH occupation fraction as the fraction of halos in a mass bin that host at least one MBH.  In Figure \ref{OccupationFractionz5}, we show the resulting MBH occupation fraction at $z=5$ for each of our recoil models.  Incorporation of recoil and spin into the MBH merger tree does yield a reduction in MBH occupation fraction.  The `no spin' recoil shows a slight decrease in the MBH occupation fraction in the halo mass range $\sim 3 \times 10^8 - 10^{10}$ M$_{\odot}$.  The spin configurations associated with larger recoil velocities exhibit a larger overall decrease in MBH occupation over a slightly larger range of halo masses, $\sim 3 \times 10^8 - 3\times 10^{10}$ M$_{\odot}$.  However, the differences between the predicted occupation fractions for the different spin recoil models are not larger than the errors associated with the measurements, and therefore not statistically significant.  The error bars on the fiducial `no kick' model and the `no spin' model represent the standard error of the mean, since these models require no randomly-generated component in the calculation of the recoil velocity.  The occupation fractions of the three remaining spin-recoil models, which require randomly selected components to compute recoil velocities, are averaged over 100 Monte Carlo trials.  Note that the error bars on these occupation fractions therefore encompass both standard error and random error.

When gravitational recoil is considered, multiple bursts of MBH formation in these simulations effectively refill host halos from which merged MBHs were previously ejected. These refill events prevent the depletion of MBH host halos, and thereby prevent a drastic reduction of the MBH occupation fraction.  As discussed in \citet{Dunn18}, our MBH formation recipe does not prevent more than one MBH seed from forming per halo, though we do take measures to prevent the spurious overproduction of MBH seeds in a single time step within a volume enclosing two softening lengths.  As a result, a single halo can experience multiple sequential episodes MBH formation.  Without gravitational recoil, as represented by our `no kick' model, MBHs that form in the same host halo merge to form a more massive seed by $z=5$.

\begin{figure}[h]
\centering
\includegraphics[width=\columnwidth]{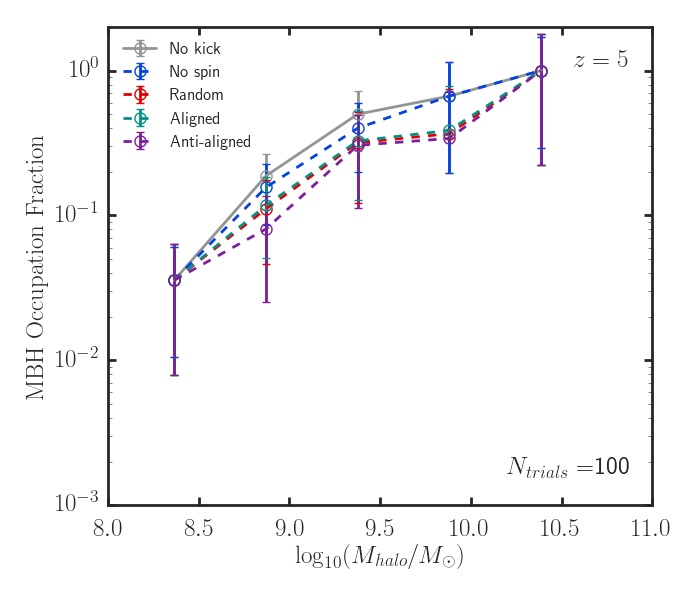}
\caption{\small{\textsc{occupation fractions}  MBH occupation fractions for different spin distributions at $z=5$.  The MBH occupation fractions computed for the random, aligned, and anti-aligned spin configurations are averaged over $100$ trials.  The error bars on the `no kick' and `no spin' models represent the standard error of the mean, whereas the error bars for the remaining models represent contributions from both standard error and random error derived from the Monte Carlo trials.  While the MBH occupation fraction appears smaller for the spin distributions associated with larger recoil velocities (random, aligned, and anti-aligned spins), these differences are not statistically significant.  The MBH occupation fraction remains similar to its fiducial value even with gravitational recoil ejections because multiple epochs of MBH formation can refill an empty host halo.}}
\label{OccupationFractionz5}
\end{figure}

%% FIGURE 2 -- VELOCITIES
The changes in the shape of the occupation fraction in Figure \ref{OccupationFractionz5} are determined by the fraction of mergers with recoil velocities greater than the host halo escape velocity.  In Figure \ref{velocityHist}, we compare the distributions of recoil velocities generated by different recoil models to the distribution of host halo escape velocities.  The distribution of host halo escape velocities is shown in solid grey, with a mean and median escape velocities of 43 km s$^{-1}$ and 47 km s$^{-1}$, respectively. The recoil velocities largely tend to be either orders of magnitude smaller or larger than the the escape velocities of the host halos, which themselves only span a single order of magnitude.  The random spin and anti-aligned spin recoil models are generally associated with kick velocities greater than or equal to the escape velocities of the host halos, nearly always resulting an ejection.  Otherwise, the `no spin' and aligned spin recoil models generate velocities spread over a larger range of magnitudes, and are less likely to be ejected.  Since halos are still quite small at these high redshifts, the recoil velocities can be comparable to, or significantly larger than, the halo escape velocities.  In turn, it is important to consider the effects of recoil in our simulations because our assumptions about MBH escape fractions and merged masses will leave an imprint on the predicted observables.

\begin{figure}[h]
  \centering
    \includegraphics[width=\columnwidth]{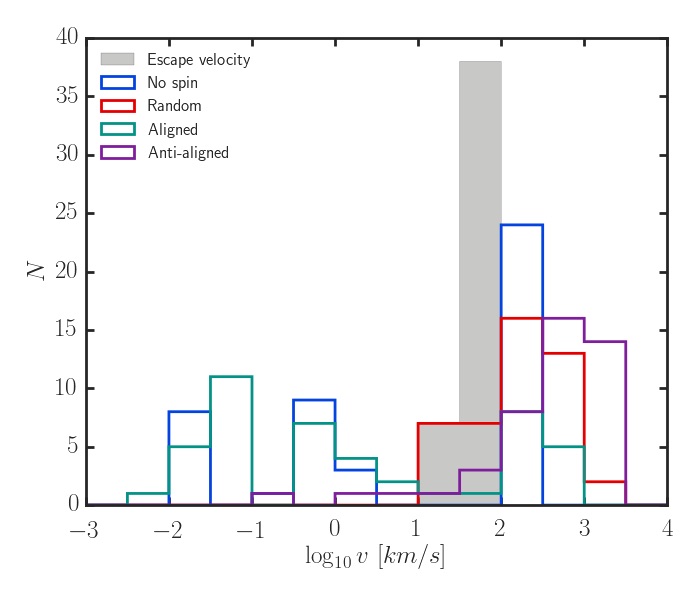}
    \caption{\small{\textsc{velocity distributions} We compare the distribution of escape velocities of host halos, shown in grey, with the distributions of recoil velocities in each of the spin configuration models, shown in their corresponding colors (see legend).  In general, the recoil velocities are either much smaller or much larger than the escape velocities, which only span one order of magnitude.}}
    \label{velocityHist}
\end{figure}

%% FIGURE 3 -- RECOIL TO ESCAPE VELOCITY RATIOS
While the focus of this work is recoil events that fully eject MBHs from the host halo, recoil events that cause MBHs to reach velocities that are a significant fraction of the host halo escape velocity are likely also important to the mass assembly of the MBH.  Recoil events for which the recoil velocity is larger than the host halo escape velocity are will eject the MBH from the halo.  In scenarios where the recoil velocity is a significant fraction of the host halo escape velocity, the MBH is likely kicked out of the central region of the halo and then `sloshes' back towards the center of the halo.  This motion may play an important role in regulating the growth of the black hole.  As the MBH wanders throughout the host halo, its motion can prevent capture into another binary system \citep{Guedes11}, and stifle growth through limited gas accretion \citep[e.g.][]{Blecha08}.  Since these simulations do not resolve dynamical friction, we are not able to fully model the orbit of such MBHs. However, we can note that recoil events in the `no spin' and aligned-spin models are unlikely to be kicked far from their original locations due to a merger, whereas recoil events in the random spin and anti-aligned spin models are more likely to be displaced from their original locations, due to their large kick velocities.

%% FIGURE 4 -- MERGER RATE
One commonly accepted idea about MBH-MBH mergers is that they are preceded and triggered by galaxy mergers.  In our simulations, MBH-MBH mergers also happen in a single halo as a direct result of the formation of multiple MBH seeds in that halo.  In Figure \ref{mergerRate}, we show the number of mergers as a function of redshift in each of our models.  We emphasize here that this figure shows merger histories, not merger rates.  Since this work uses zoom-in simulations, we do not generate a statistical sample that would be necessary to calculate the rate of MBH mergers that may be observable with LISA.  The redshift distribution of MBH-MBH mergers closely mirrors that of MBH formation, shown in black.  The number of black hole mergers in the original simulation, with no recoil model, is shown by the filled gray bars.  Each of the four gravitational recoil models decrease the number of black hole mergers in the assembly history of this galaxy.  The MBH-MBH mergers that we recover in this simulation are a direct result of the formation of multiple MBH seeds within small volumes of space and windows of time.  This indicates that some of the mergers detected by LISA may be of black holes that recently formed in a single halo and are still very close to the original seed mass.

\begin{figure}[h]
    \centering
      \includegraphics[width=\columnwidth]{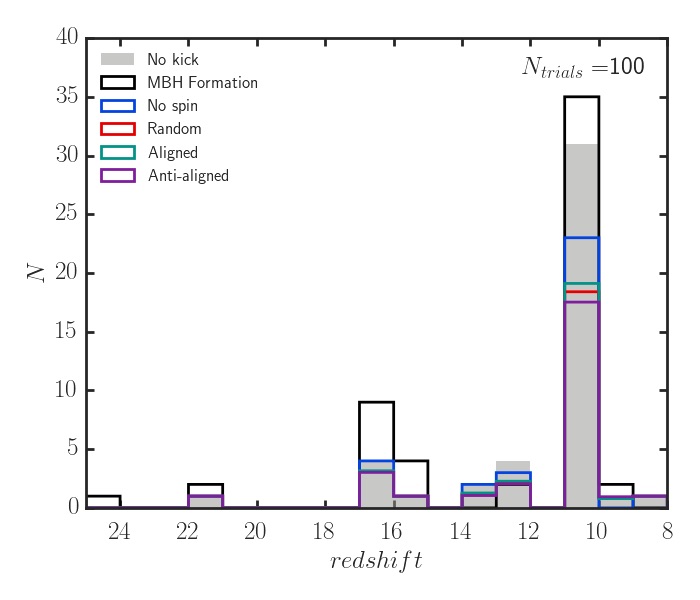}
      \caption{\small{\textsc{MBH-MBH merger rates} Comparison of the frequency of MBH mergers for different spin distributions to the MBH formation rate.  The MBH formation history is shown in black, and the merger history for the original simulation with no recoil model is shown by the filled gray bars.  While recoil models that result in larger kick velocities suppress the number of MBH-MBH mergers, all of the models show heightened merger rates during episodes of MBH formation.  Merger histories computed for the random, aligned, and anti-aligned spin configurations are averaged over $100$ trials.}}
      \label{mergerRate}
\end{figure}

%% FIGURE 5 -- HALO-BH MASS
Gravitational recoil events likely contribute to the intrinsic scatter in MBH-host scaling relations \citep{Libeskind06,Volonteri07,Devecchi09b}.  In Figure \ref{haloVbhMass}, we demonstrate the effect of different spin recoil models on the halo mass$-$MBH mass relation.  While a halo may still host an MBH seed at $z=5$ when gravitational recoil is considered, multiple episodes of ejection and replenishment affects the mass of that seed.  The incorporation of a recoil model can decrease the mean final MBH mass by as much as an order of magnitude.  The random spin and anti-aligned spin models typically show the largest decrease in mean final MBH mass.  Hosts that show no change in the mean final MBH mass do not experience any MBH-MBH mergers.  The mean final MBH masses in these cases are equal to the initial MBH seed mass.  Here we demonstrate two pathways for a halo in this simulation to to retain a central black hole with a mass resembling that of the original seed down to $z=5$.  In one scenario, a host halo may form only a single seed, and experience no events that feed this seed via either mergers with other black holes or accretion.  Alternatively, a host may form multiple seeds at high redshift, but repeatedly eject the merged remnants from its shallow potential well. 

\begin{figure}[h]
    \centering
      \includegraphics[width=\columnwidth]{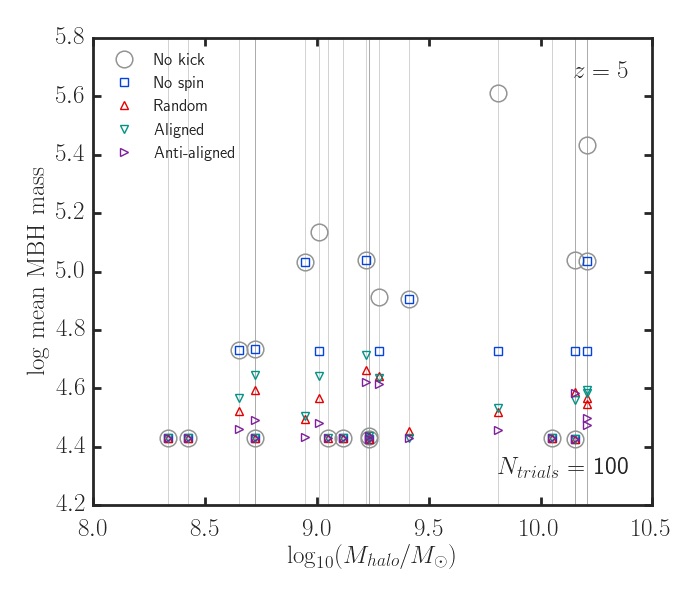}
      \caption{\small{\textsc{halo mass versus black hole mass} Each colored shape shows the mean final MBH masses in the different spin-recoil models.  Grey vertical lines mark the mass of each host halo and serve to guide the eye.  Vertical displacement between the fiducial `no kick' mass and the final MBH masses associated with other models indicate that an MBH's assembly history was modified by at least one episode of merging, ejection, and refill.  At $z=5$, the incorporation of recoil velocity and different spin models decrease the average MBH mass.  The random and anti-aligned spin configurations yield lower final average MBH masses.  Black hole masses computed for the random, aligned, and anti-aligned spin configurations are averaged over $100$ trials.}}
      \label{haloVbhMass}
\end{figure}

Similarly, in Figure \ref{stellarVbhMass}, we demonstrate the effect of different spin recoil models on the stellar mass$-$MBH mass relation.  We note here that we multiply the stellar masses by 0.6 to convert simulated masses to observed masses in order to reproduce the stellar mass-halo mass relation as described in \citet{Munshi13}.  For comparison, we overplot the $z=0$ $M_{\rm BH}-M_{\rm stellar}$ relations for local galaxies with stellar masses in the range $10^8 - 10^{12} M_{\odot}$ provided by \citet{Reines15}.  Gravitational recoil can be a significant source of scatter in these scaling relations, potentially up to an order of magnitude. This effect can skew $M_{\rm BH}-M_{\rm stellar}$ values below the expected relations, by indirectly increasing central star formation.
 
%% FIGURE 6 -- STELLAR-BH MASS
\begin{figure}[h]
    \centering
      \includegraphics[width=\columnwidth]{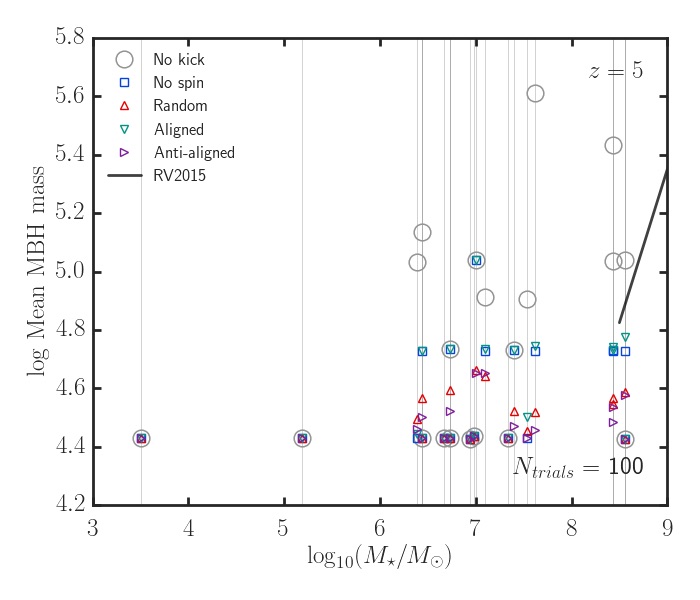}
      \caption{\small{\textsc{stellar mass versus black hole mass} This figure is constructed similarly to Figure \ref{haloVbhMass}, but instead shows black hole mass as a function of stellar mass.  The random and anti-aligned spin configurations yield lower final average MBH masses.  Grey vertical lines mark the stellar mass of each host and serve to guide the eye.  Black hole masses computed for the random, aligned, and anti-aligned spin configurations are averaged over $100$ trials.  For comparison, we include the $z=0$ relation described by \citet{Reines15} for halos with stellar masses in the range $10^8 - 10^{12} M_{\odot}$.}}
      \label{stellarVbhMass}
\end{figure}

%% FIGURE 7 -- WATERFALL
Mergers are plentiful in the assembly history of this Milky Way-type galaxy.  Even if MBH mergers are electromagnetically dark, they are ideally suited for detection through gravitational waves.  LISA is optimally designed to detect MBH mergers in the low-frequency regime, ranging from $10^4$ to $10^7$M$_{\odot}$ in mass and out to redshifts larger than $z\sim$ 20 \citep{LISA17}.  In Figure \ref{waterfall}, we show the distribution of MBH merger masses and redshifts generated in this work against LISA $S/N$ ratio contours in a `waterfall' figure.  Each black point marks a merger in our `no spin' semi-analytic recoil model against the $S/N$ contours that are a function of the redshift and combined total mass of the merging MBHs.  These $S/N$ values are approximate, as the rainbow contours are generated for 1:4 mass ratio mergers, but the most common mergers in these simulations have 1:1 and 1:2 mass ratios.  These mergers occur throughout the epoch of MBH formation, and the cluster of events at $z\sim11$ coincides with the peak of MBH formation. LISA detections of MBH-MBH mergers will help constrain MBH formation scenarios, masses, and redshifts \citep{Colpi19}.

The next generation of space-based observatories will offer some of the first observational constraints of the formation and early growth of massive black hole seeds.  The planned Lynx mission will have the sensitivity to detect objects at the low-luminosity and high-redshift ends of the quasar luminosity function.  The James Webb Space Telescope (JWST) could potentially differentiate between massive black hole seed formation mechanisms \citep[e.g.,][]{Natarajan17}.  LISA will constrain the population of black holes with masses $10^4-10^7M_{\odot}$, offering the first observational constraints of these electro-magnetically dark high-redshift objects.  Assumptions about MBH merger fractions and recoil fractions, even physically motivated assumptions, will leave an imprint on predictions of the observable signatures of these objects.  Predictions of the unresolved X-ray background, gravitational wave events, and the quasar luminosity function (especially the low-luminosity end) \citep[e.g.,][]{Manti17,Ricarte18}, for example, are directly affected by the assumed merger and recoil fractions.  Any work that attempts to disentangle the relationship between black hole formation and AGN observables must also address the physics of black hole mergers and gravitational recoil.

\begin{figure}[h]
  \centering
    \includegraphics[width=\columnwidth]{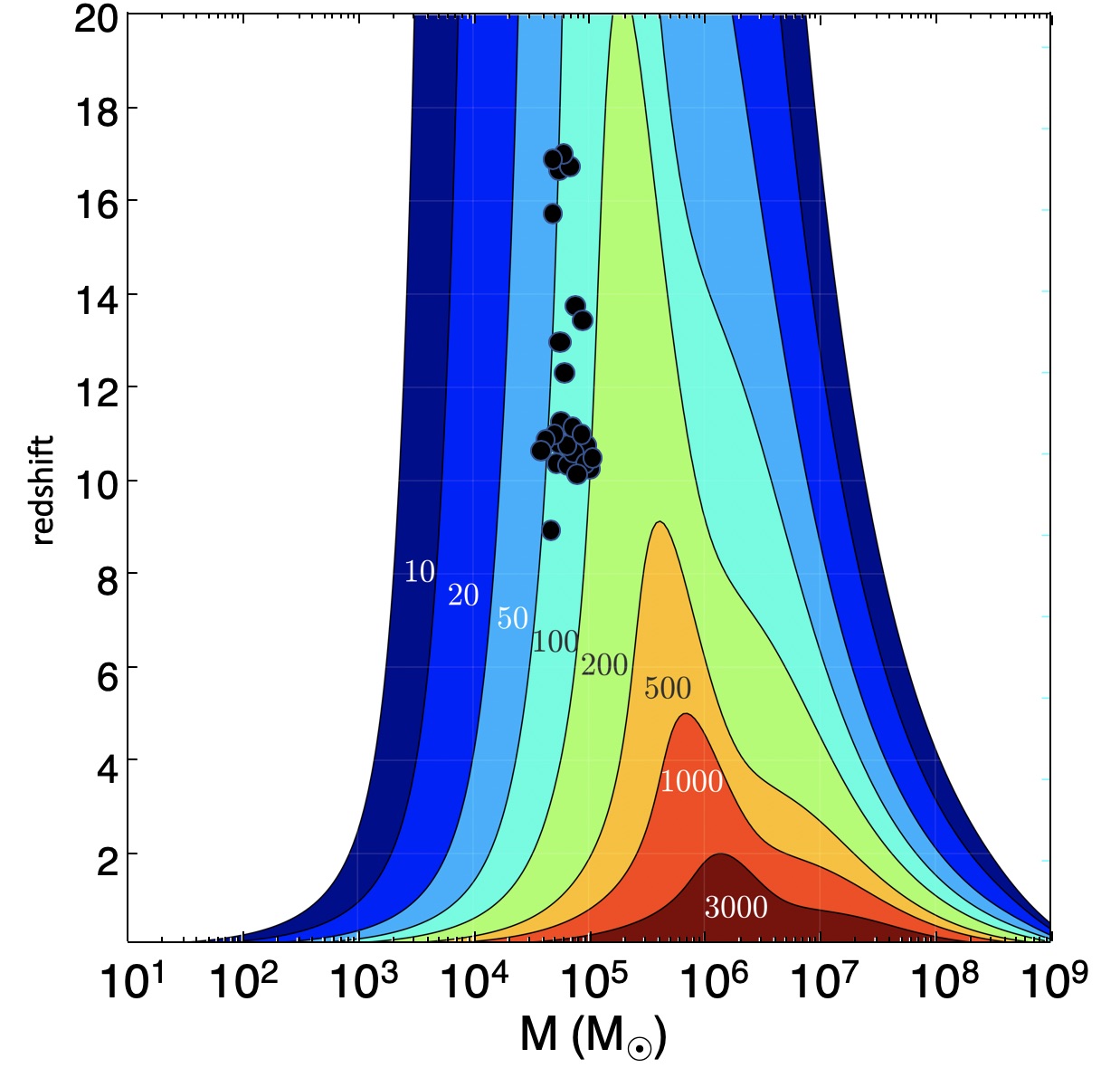}
    \caption{\small{\textsc{lisa signal} Mergers are plentiful in the assembly history of this galaxy.  The black points in this figure represent the MBH mergers that occur during the assembly of this Milky Way-type halo.  The colored contours represent the $S/N$ ratios with which LISA would detect mergers with a mass ratio of 0.25.  This sample of synthetic LISA source signals is drawn from the `no spin' recoil model.  The large cluster of sources in the redshift range $z\sim 10-12$ represents the spike in MBH-MBH mergers that results from a simultaneous spike in MBH formation.  The merger that occurs at $z=22$ is not shown.}}
    \label{waterfall}
\end{figure}

%%%%%%%%%%%%%%%%%%%%%%%%%%%%%%%%%
\section{Summary} \label{Summary}
%%%%%%%%%%%%%%%%%%%%%%%%%%%%%%%%%
We studied the role of gravitational recoil in the mass assembly of massive black hole seeds using cosmological zoom-in simulations and a semi-analytic model for gravitational recoil velocities.  Our results underline that gravitational wave recoil stifles the early growth of the seed MBHs, and alters the assembly pathways of MBHs and their host galaxies.  In environments where multiple MBHs can form in bursts, gravitational recoil can prevent the rapid growth of black holes by mergers, as these mergers typically eject the remnant.  However, these same bursts of black hole formation also allow an empty halo to be refilled with new MBH seeds.  The restocking of a proto-galactic host is reflected in the statistical agreement of MBH occupation fractions associated with different recoil models.  The masses of the final MBHs, however, may be significantly reduced, and more closely resemble the initial seed mass.  The differences in final MBH masses may inject scatter into observable scaling relations and provide tension with these relations.

MBH mergers are likely an important component of MBH-galaxy co-evolution.  Even if an MBH-MBH merger is not ejected from its host, the recoil velocity may still be large enough to displace the MBH from the center of its host \citep{Blecha08,Guedes11} or even force the MBH to wander the outskirts of the host galaxy or dark matter halo \citep{HolleyBockelmann08,Bellovary10}.  As the MBH wends its way toward the center via dynamical friction, accretion is likely minimal \citep{vanWassenhove10}, but if it encounters a pocket of gas with low relative velocity (e.g., at apocenter), it can outshine its host as an off-center AGN \citep{Blecha16,Comerford15} as seen in NGC 3115 \citep{Menezes14}.  This trajectory through the host allows the merged MBH to accrete, modifying the object's mass and spin, and possibly allow it to avoid capture into another binary system \citep{Guedes11}. 
 
MBH-MBH mergers are ideal LISA candidates, particularly mergers of MBH seeds at redshifts 10$-$20 such as those discussed in this work.  LISA observations of gravitational wave emissions from these mergers will likely provide some of the first insights into the lives of the first black holes.  While gas accretion erases clues about MBH formation long before these black holes are electromagnetically observable, gravitational waves from pre-reionization merger events will provide a more direct probe of the seed black hole population.  These observations will characterize the number density, masses, and redshifts of black hole mergers, and have the potential to help constrain seed formation scenarios.

In future work, it will be important to study this phenomenon with a suite of galaxies at different masses and assembly histories.  If ejections of merged binary black holes are common, gravitational recoil may be a serious impediment to the rapid accumulation of black hole mass necessary to create both high-$z$ quasars and low-$z$ black hole-host scaling relations.

\acknowledgements
JMB is grateful for support from NSF award AST-1812642.

%\clearpage
%\pagebreak
\bibliography{recoilPaper}
%\nocite{*}
%%\clearpage

\end{document}